\documentstyle[12pt]{article}

\newcommand{\be}{\begin{equation}}
\newcommand{\ee}{\end{equation}}
\newcommand{\bea}{\begin{eqnarray}}
\newcommand{\eea}{\end{eqnarray}}
\newcommand{\beaa}{\begin{eqnarray*}}
\newcommand{\eeaa}{\end{eqnarray*}}

\newcommand{\J}{{\cal J}}

\newcommand{\BB}{{{\rm I} \kern -2pt \rlap {\rm B} \kern +8pt}}

\advance\textheight by \topskip
\makeatletter

\@addtoreset{equation}{section}
\def\section{\@startsection {section}{1}{\z@}{-3.5ex plus -1ex minus
 -.2ex}{2.3ex plus .2ex}{\large\bf\centering}}
\def\subsection{\@startsection{subsection}{2}{\z@}{-3.25ex plus -1ex minus -.2ex}{1.5ex plus .2ex}{\bf}}
\def\subsubsection{\@startsection{subsubsection}{3}{\z@}{-3.25ex plus -1ex minus -.2ex}{1.5ex plus .2ex}{\sl}}
\makeatother

\begin{document}

\baselineskip 18pt \parindent 12pt \parskip 10pt

\begin{titlepage}

\begin{center}
{\Large {\bf Lax pair and Darboux transformation of noncommutative
$U(N)$ principal chiral model }}\\\vspace{1.5in} {\large
U. Saleem\footnote{%
usman\_physics@yahoo.com}  and M. Hassan \footnote{%
mhassan@physics.pu.edu.pk} }\vspace{0.15in}

{\small{\it Department of Physics,\\ University of the Punjab,\\
Quaid-e-Azam Campus,\\Lahore-54590, Pakistan.}}
\end{center}

\vspace{1cm}
\begin{abstract}
\noindent We present a noncommutative generalization of Lax
formalism of $U(N)$ principal chiral model in terms of a
one-parameter family of flat connections. The Lax formalism is
further used to derive a set of parametric noncommutative
B\"{a}cklund transformation and an infinite set of conserved
quantities. From the Lax pair, we derive a noncommutative version
of the Darboux transformation of the model.
\end{abstract}
\vspace{1cm} PACS: 11.10.Nx, 02.30.Ik
\end{titlepage}

\section{Introduction}

During the last decade, there has been an increasing interest in
the study of noncommutative field theories (nc-FTs) due to their
relation to string theory, perturbative dynamics, quantum Hall
effect etc \cite{N1}-\cite{moh}. The noncommutative field theories
can be constructed in different settings. One way of this
construction is through Moyal deformation product or
$\star$-product(Moyal product) \cite{moyal}.  A simple
noncommutative field theory (nc-FT) can be obtained by replacing
the product of fields by their $\star$-product. The
noncommutativity of coordinates of the Euclidean space $R^D$ is
defined as
\[
\left[ x^{\mu},x^{\nu}\right] =\mbox{i}\theta ^{\mu \nu},
\]
where $\theta ^{\mu \nu}$ is a second rank antisymmetric real
constant tensor known as deformation parameter. This has been
shown that in general, noncommutativity of time variables leads
to non-unitarity and affects the causality of the theory \cite{se, go}. The $%
\star$-star product of two functions in noncommutative Euclidean
spaces is given by
\[
\left( f\star g\right) (x)=f(x)g(x)+\frac{\mbox{i}\theta ^{\mu
\nu}}{2}\partial _{\mu}f(x)\partial _{\nu}g(x)+\vartheta (\theta
^{2}), \]where $\partial _{\mu}=\frac{\partial }{\partial
x^{\mu}}$. These nc-FTs reduce to the ordinary or commutative
field theories (FTs) as the deformation parameter reduces to zero.
There is also an increasing interest in the noncommutative
extension of integrable field theories (nc-IFTs)
\cite{Fur}-\cite{usman}. Some times the noncommutativity breaks
the integrability of a theory, however there are some examples in
which integrability of a field theory is maintained \cite{Marco1}.
In reference \cite{Marco1}, a noncommutative extension of
$U(N)$-principal chiral model (nc-PCM) has been presented and it
is concluded that this noncommutative extension gives no extra
constraints for the theory to be integrable. The non-local
conserved quantities of nc-PCM have also been derived using the
iterative method of Br\'{e}zin-Itzykson-Zinn-Zuber (BIZZ)
\cite{BIZZ} but no effort has been made so far to study the Lax
formalism of the nc-PCM and to derive conserved quantities and
Darboux transformation from it.

In this paper we present a Lax formalism of one-parameter family
of transformations on solutions of noncommutative $U(N)$ principal
chiral model. The Lax formalism further gives a set of parametric
noncommutative B\"{a}cklund Transformation (nc-BT) and a set of
Riccati equations. The Lax formalism can be used to derive a
series of conserved quantities. The Lax formalism is further used
to develop the noncommutative version of Darboux transformation
for the nc-PCM. We expand the Noether currents in power series in
deformation parameter and obtain zeroth and first order equations
of motion and the conserved quantities.

\section{Noncommutative principal chiral model}\label%
{nc-pcm}

The action of $U(N)$ for the nc-PCM is defined by \footnote{The
nc-PCM can also be obtained by dimensional reduction of
noncommutative anti-self dual Yang-Mills equations in four
dimensions \cite{Takasaki}.}
\begin{equation} {\cal
S^\star}= {\frac{1 }{2}}\int d^2 x {\rm Tr}( \partial_+
g^{-1}\star
\partial_- g ),  \label{action2}
\end{equation}
with constraints on the fields $g(x^{+},x^{-})$
\[
g^{-1}(x^{+},\,x^{-})\,\star
g(x^{+},\,x^{-})=g(x^{+},\,x^{-})\star g^{-1}(x^{+},\,x^{-})=1,
\]
where  $g(x^{+},x^{-})\in U(N)$\footnote{Here we have taken the
global symmetry group as $U(N)$, which has a simple noncommutative
extension. There is no noncommutative $SU(N)$ because
det$(g_{1}\star g_{2})\neq$ det$(g_{1})\star$ det$(g_{2})$. Also
for any $X,Y \in su(N)$ Lie algebra of the Lie group $SU(N)$, the
commutator $X \star Y-Y \star X$ is not traceless. The
noncommutative extensions of $SO(N)$ and $US_{P}(N)$ have been
constructed in \cite{R1}-\cite{R3} but the construction is a bit
involved.} and $g^{-1}(x^{+},\,x^{-})$ stands for an inverse with
respect to the $\star$-product \footnote{Our conventions are such
that the two-dimensional coordinates are related by
$x^{\pm}=\frac{1}{2}(x^{0} \pm ix^{1})$ and
$\partial_{\pm}=\frac{1}{2}(\partial_{0} \pm i\partial_{1})$.}.
The $U(N)$-valued field $g(x^{+},x^{-})$ is defined as
\[
g(x^{+},x^{-})\equiv e_{\star}^{\mbox{i}
\pi_{a}T^{a}}=1+\mbox{i}\pi_{a}T^{a}+\frac{1}{2}(\mbox{i}\pi_{a}T^{a})_{\star}^2+\dots,
\]
where $\pi_{a}$ is in the Lie algebra $u(N)$ of the Lie group
$U(N)$ and $T^{a}$, $a=1,2,3\dots, N^2$ are hermitian matrices
with the normalization \rm Tr$(T^{a} T^{b})=-\delta^{ab}$ and are
the generators of $U(N)$ in the fundamental representation
satisfying the algebra
\[
[T^{a},T^{b}]=\mbox{i}f^{abc}T^{c},
\]
where $f^{abc}$ are the structure constants of the Lie algebra
$u(N)$. For any $X\in u(N)$, we write $X=X^{a}T^{a}$ and $X^{a}=-$
\rm Tr$(T^{a}X)$.

The action (\ref{action2}) is invariant under a global continuous
symmetry
\[
U_{L}(N)\times U_{R}(N): \qquad \qquad g(x^{+},\,x^{-})\mapsto u
\star g \star v^{-1}.
\]
The associated Noether conserved currents of nc-PCM are
\begin{eqnarray*}
j_{\pm }^{\star }\,^{R} =-g^{-1}\star \partial _{\pm }g,\qquad
j_{\pm }^{\star }\,^{L} =\partial _{\pm }g\star g^{-1},
\end{eqnarray*}
which take values in the Lie algebra $u(N)$, so that one can
decompose the currents into components
$j_{\pm}^{\star}(x^{+},x^{-})=j_{\pm}^{\star a
}(x^{+},x^{-})T^{a}$. The equation of motion following from
(\ref{action2}) corresponds to conservation of these currents. The
left and right currents satisfy the following conservation
equation
\begin{eqnarray}
\partial _{-}j_{+}^{\star }\,+\partial _{+}j_{-}^{\star }\, &=&0.  \label{m1}
\end{eqnarray}
The currents also obey the zero-curvature condition
\begin{eqnarray}
\partial _{-}j_{+}^{\star }\,-\partial _{+}j_{-}^{\star }\,+\left[
j_{+}^{\star },\,j_{-}^{\star }\right] _{\star } &=&0,  \label{zero1}
\end{eqnarray}
where $\left[ j_{+}^{\star },\,j_{-}^{\star }\right] _{\star
}=\,\,j_{+}^{\star }\,\,\,\star
\,\,j_{-}^{\star }-j_{-}^{\star }\,\star \,\,j_{+}^{\star }$. The equations (%
\ref{m1}) and (\ref{zero1}) can also be expressed as
\begin{equation}
\partial _{-}j_{+}^{\star }\,=-\partial _{+}j_{-}^{\star }\,=-\frac{1}{2}%
\left[ j_{+}^{\star },\,j_{-}^{\star }\right] _{\star }.  \label{mzn}
\end{equation}
The equations (\ref{m1}), (\ref{zero1}) and (\ref{mzn}) holds for both $j_{\pm}^{\star L}$ and $%
j_{\pm}^{\star R}$.

\section{Lax pair and conserved quantities of nc-PCM}\label{tran}
In order to develop a Lax formalism and construct infinitely many
conserved quantities for the nc-PCM, we define a one-parameter
family of transformations on field $g(x^{+},\,x^{-}) $ in
noncommutative space as
\[
g\rightarrow g^{(\gamma )}=u^{(\gamma )}\star \,g\star \,\,v^{(\gamma )-1},
\]
where $\gamma $ is a parameter and $u^{(\gamma )}$, $v^{(\gamma
)}$ are
matrices belonging to $U(N)$. We choose the boundary values $%
u^{(1)}=1,v^{(1)}=1$ or $g^{(1)}=g$. The matrices $u^{(\gamma )}$
and $v^{(\gamma )}$ satisfy the following set of linear equations
\begin{eqnarray}
\partial _{\pm}{u^{(\gamma )}} &=&{\frac{1}{2}}(1-\gamma ^{\mp1}) j_{\pm}^{\star
L}\,\star {u^{(\gamma )},}  \label{l1} \\
\partial _{\pm}{v^{(\gamma)}} &=&{\frac{1}{2}}(1-\gamma
^{\mp1}) j_{\pm}^{\star R}\,\star {v^{(\gamma )}}.  \label{l3}
\end{eqnarray}
In what follows, we shall consider right hand currents and drop
the superscript $R$ on the current to simply write $j_{\pm
}^{\star R}=j_{\pm }^{\star}$. The compatibility condition of the
linear system (\ref{l3}) is given by
\begin{eqnarray*}
\left\{ (1-\gamma ^{-1})\partial _{-}j_{+}^{\star}-(1-\gamma
)\,\partial _{+}j_{-}^{\star}+(1-{{\frac{1}{2}}}(\gamma +\gamma
^{-1}))[j_{+}^{\star}\,,\,j_{-}^{\star}]_{\star }\right\} \star
{v^{(\gamma )}}\, &=&\,0.
\end{eqnarray*}
Under the one-parameter family of transformation, the Noether
conserved currents transform as
\begin{eqnarray*}
j_{\pm}^{\star} &\mapsto &j_{\pm}^{\star (\gamma )}\,=\,\gamma ^{\mp1}v{%
^{(\gamma )-1}\star }j_{\pm}^{\star }\star {v^{(\gamma )}}.
\end{eqnarray*}
The one-parameter family of conserved currents $j_{\pm}^{\star (\gamma)}$ in noncommutative space for any value of $\gamma $%
: $\partial _{+}j_{-}^{\star \,(\gamma )}+\partial
_{-}j_{+}^{\star (\gamma )}=0$. The linear system (\ref{l3}) can
be written as
\begin{equation}
\partial _{\pm }v(x^{+}, x^{-};\lambda )=A_{\pm }^{\star (\lambda )}\star
v(x^{+}, x^{-};\lambda ),  \label{nclinear}
\end{equation}
where the noncommutative fields $A_{\pm }^{\star (\lambda )}$ are
given by
\[
A_{\pm }^{\star (\lambda )}=\mp {\frac{\lambda }{1\mp \lambda }}j_{\pm
}^{\star }.
\]
The parameter $\lambda $ is spectral parameter and is related to parameter $%
\gamma $ by $\lambda =\frac{1-\gamma }{1+\gamma }$. The
compatibility condition of the linear system (\ref{nclinear}) is
the $\star$-zero-curvature condition
\begin{equation}
\left[ \partial _{+}-A_{+}^{\star (\lambda )},\partial _{-}-A_{-}^{\star
(\lambda )}\right] _{\star }\equiv \partial _{-}A_{+}^{\star (\lambda
)}-\partial _{+}A_{-}^{\star (\lambda )}+\left[ A_{+}^{\star (\lambda
)},A_{-}^{\star (\lambda )}\right] _{\star }=0.  \label{zero12}
\end{equation}
We have defined a one-parameter family of connections
$A_{\pm}^{\star (\lambda)}$ which are flat. The Lax operators can
now be defined as
\begin{equation}
L_{\pm }^{\star (\lambda )}=\partial _{\pm }-A_{\pm }^{\star (\lambda )},
\label{Lax1}
\end{equation}
which obey the following equations
\begin{equation}
\partial _{\mp }L_{\pm }^{\star (\lambda )}=[A_{\mp }^{\star (\lambda
)}\,,L_{\pm }^{\star (\lambda )}]_{\star }.  \label{Lax2}
\end{equation}
The associated linear system (\ref{nclinear}) can be re-expressed
as
\begin{equation}
\partial _{0}v(x^{0}, x^{1};\lambda )=A_{0}^{\star (\lambda )}\star v(x^{0}, x^{1};\lambda
),\qquad \partial _{1}v(x^{0}, x^{1};\lambda )=A_{1}^{\star
(\lambda )}\star v(x^{0}, x^{1};\lambda ),\label{dual1}
\end{equation}
with the noncommutative connection fields $A_{0}^{\star (\lambda
)}$ and $A_{1}^{\star (\lambda )}$ given by
\[
A_{0}^{\star (\lambda )}=-{\frac{\lambda }{{1-\lambda ^{2}}}}\left(
j_{1}^{\star }+\lambda j_{0}^{\star }\right) ,\qquad A_{1}^{\star (\lambda
)}=-{\frac{\lambda }{{1-\lambda ^{2}}}}\left( j_{0}^{\star }+\lambda
j_{1}^{\star }\right) .
\]
The compatibility condition of the linear system (\ref{dual1}) is
again the $\star$-zero-curvature condition for the fields
$A_{0}^{\star (\lambda )}$ and $A_{1}^{\star (\lambda )}$
\[
\left[ \partial _{0}-A_{0}^{\star (\lambda )},\partial _{1}-A_{1}^{\star
(\lambda )}\right] _{\star }\equiv \partial _{1}A_{0}^{\star (\lambda
)}-\partial _{0}A_{1}^{\star (\lambda )}+\left[ A_{0}^{\star (\lambda
)},A_{1}^{\star (\lambda )}\right] _{\star }=0.
\]
The Lax operator is defined as
\[
L_{1}^{\star (\lambda )}=\partial _{1}-A_{1}^{\star (\lambda )},
\]
which obeys the Lax equation
\[
\partial _{0}L_{1}^{\star (\lambda )}=[A_{0}^{\star (\lambda )},L_{1}^{\star
(\lambda )}]_{\star }.
\]
This equation gives the $x^{0}$-evolution of the operator
$L_{1}^{\star (\lambda )}$ and is equivalent to an isospectral
eigenvalue problem. We have been able to show that the existence
of a one-parameter family of transformations and Lax formalism of
PCM can be generalized to nc-PCM without any constraints. This
works straight forwardly as it does in the commutative case. The
one-parameter family of transformations thus gives rise to an
infinite number of conserved quantities and the Darboux
transformation of generating solution of nc-PCM.

\subsection{Local conserved quantities} It is straight forward to derive an
infinite set of local\footnote{The term "local" in our discussion
refers to its standard meaning. The conserved densities depend
upon fields and their derivatives but not on their integrals. The
intrinsic non-locality of the Moyal deformation products appearing
due to the presence of derivatives to infinite order, persists in
all our discussions. These conserved quantities are in fact
deformed local conserved quantities carrying with them the
intrinsic non-locality due to noncommutativity. Moreover, the
leading terms in the perturbative expansion in $\theta$, are
local.} conserved quantities from the equation (\ref{mzn})
\begin{eqnarray}
\partial _{\mp }{\rm Tr}\left( j_{\pm }^{\star }\right) ^{n}=0\label{localseries},
\end{eqnarray} where $n$ is an integer and the first case $n=2$
corresponds to the conservation of the energy momentum tensor.
These conservation laws are associated with the totally symmetric
invariant tensors of the Lie algebra $u(N)$ and the integers $n$
turn out to be the exponents of $u(N)$. This is exactly what
happens in the commutative case and these conserved quantities are
shown to be in involution with each other \cite{Gold}, in the
commutative case.

We can also derive the local conserved quantities of
nc-PCM from the linear system (\ref{nclinear}) via noncommutative B\"{a}%
cklund transformation (nc-BT) and Riccati equations. The linear
system (\ref {nclinear}) reduces to the following set of
noncommutative B\"{a}cklund transformation (nc-BT)
\begin{eqnarray}
\pm\partial _{\pm}(g^{-1}\star \tilde{g})
&=&\tilde{j}_{\pm}^{\star }-j_{\pm}^{\star },  \label{BT1}
\end{eqnarray}
with constraint $g^{-1}\star \tilde{g}+\tilde{g}^{-1}\star
g=2\lambda ^{-1}I$, where $\lambda$ is a real parameter, $g$ and
$\tilde{g}$ are solutions of equation of motion. The nc-BT given
by equation (\ref{BT1}) further gives rise to a set of compatible
noncommutative Riccati equations
\begin{eqnarray}
\partial _{\pm}\Gamma(\lambda ) &=&-\frac{\lambda }{2(1\mp\lambda )}\left(
j_{\pm}^{\star }+\Gamma(\lambda )\star j_{\pm}^{\star }\star
\Gamma(\lambda )-2\lambda ^{-1}j_{\pm}^{\star }\star\Gamma(\lambda
)\mp\left[ \Gamma(\lambda ),j_{\pm}^{\star }\right] _{\star
}\right) , \label{R1}
\end{eqnarray}
where $\Gamma(\lambda )=g^{-1}\star \tilde{g}$. The linearization
of the Riccati equation (\ref{R1}) gives rise to the linear system
(\ref{nclinear}). The equations (\ref {m1}), (\ref{zero1}) and
(\ref{R1}) can be used to give a series of conservation laws
\begin{eqnarray} (1+\lambda )\partial _{-}{\rm Tr}\left( \Gamma(\lambda
)\star j_{+}^{\star }\right) -(1-\lambda )\partial _{+}{\rm
Tr}\left( \Gamma(\lambda )\star j_{-}^{\star }\right)
=0\label{BTCON}.
\end{eqnarray}
Expanding $\Gamma(\lambda )$ as a power series in $\lambda $:
$\Gamma(\lambda
)=\sum_{k=0}^{\infty }\lambda ^{k}\Gamma_{k}$, one can generate $\lambda $%
-independent conservation laws. It is not easier to solve the
algebraic equations obtained recursively from (\ref{BTCON}) when
we substitute the expansion of $\Gamma(\lambda)$. The explicit
form of conservation law is therefore not quite transparent. It
is, therefore, not straight forward to relate these local
conserved quantities with the ones associated with invariant
tensors of $u(N)$.

The existence of nontrivial higher spin local conserved quantities
has important implications regarding classical and quantum
integrability of a field theory. In two dimensional quantum field
theory, their existence forces the multiparticle scattering matrix
to factorize into a product of two particle scattering matrices
and eventually to be computed exactly. The two particle $S$-matrix
satisfies the Yang-Baxter equation \cite{y1}-\cite{y3}. We expect
that the local conserved quantities in nc-PCM will also give some
important information about the complex dynamics of the model. The
higher spin  local conserved quantities of the type
(\ref{localseries}) are related to the $W$-algebra structure
appearing in certain conformal field theories. The deformed local
conserved quantities would naturally lead to the study of
deformation of $W$-algebra \cite{y4}.

\subsection{Non-local conserved quantities}

An infinite number of non-local\footnote{Here again the term
"non-local" refers to the meaning that the conserved densities
depend on fields, their derivatives and their integrals and they
also contain intrinsic non-locality of the Moyal deformation.}
conserved quantities can also be generated from the Lax formalism
of nc-PCM. We assume spatial boundary
conditions such that the currents $j^{\star (\gamma )}$ vanish as $%
x^{1}\rightarrow \pm \infty $. The equation (\ref{dual1}) implies that $%
v(x^{0},\infty ;\lambda )$ are time independent. The residual
freedom in the solution for $v(x^{0},\infty ;\lambda )$ allows us
to fix $v(x^{0},\infty ;\lambda )=1$, the unit matrix and we are
then left with $x^{0}$-independent matrix valued function
\begin{equation}
Q^{\star}(\lambda )=v(x^{0},\infty ;\lambda ).  \label{ch}
\end{equation}
Expanding $Q^{\star}(\lambda )$ as power series in $\lambda $
gives infinite number of non-local conserved quantities
\begin{equation}
Q^{\star}(\lambda )=\sum_{k=o}^{\infty }\lambda ^{k}Q^{\star(k)}\qquad ,\qquad \qquad {%
\frac{d}{dx^{0}}}Q^{\star(k)}=0.  \label{expn1}
\end{equation}
For the explicit expressions of the non-local conserved
quantities, we write (\ref {dual1}) as
\begin{equation}
v(x^{0},x^{1};{\lambda })=1-{\frac{{\lambda }}{{1-\lambda
^{2}}}}\int_{-\infty }^{x^{1}}dy\,\left( j_{0}^{\star
}(x^{0},y)-\lambda j_{1}^{\star }(x^{0},y)\right) \star
v(x^{0},y;{\lambda }).  \label{int}
\end{equation}
We expand the field $v(x^{0},x^{1};{\lambda })$ as power series in
$\lambda $ as
\begin{equation}
v(x^{0},x^{1};{\lambda })=\sum_{k=o}^{\infty }\lambda
^{k}v_{k}(x^{0},x^{1}), \label{expn}
\end{equation}
and compare the coefficients of powers of $\lambda $, we get a
series of conserved non-local currents, which upon integration
give non-local conserved quantities. The first two non-local
conserved quantities of nc-PCM are
\begin{eqnarray*}
Q^{\star (1)} &=&-\int_{-\infty }^{\infty }dy\,\,j_{0}^{\star }(x^{0},y), \\
Q^{\star (2)} &=&-\int_{-\infty }^{\infty }dyj_{1}^{\star
}(x^{0},y)+\int_{-\infty }^{\infty }dyj_{0}^{\star }(x^{0},y)\star
\int_{-\infty }^{y}dz\,j_{0}^{\star }(x^{0},z).
\end{eqnarray*}
These conserved quantities are exactly the same as obtained in
\cite{Marco1} using noncommutative iterative method of Brezin {\it
et. al. }\cite{BIZZ}{\it \ }. We now show that the procedure
outlined above is equivalent to the iterative construction of
non-local conserved quantities of nc-PCM \cite{Marco1}. From
equations (\ref{nclinear}) and (\ref {expn}), we get
\[
\partial _{\pm }\sum_{k=o}^{\infty }\lambda ^{k}v_{k}(x^{0},x^{1})=\pm D_{\pm
}\sum_{k=o}^{\infty }\lambda ^{k}v_{k}(x^{0},x^{1}),
\]
where the covariant derivatives $D_{\pm }$ are defined as
\[
D_{\pm }v^{(k)}=\partial _{\pm }v^{(k)}-j_{\pm }^{\star }\star
v^{(k)}\qquad \Rightarrow \qquad [D_{+},D_{-}]_{\star }=0.
\]
We can now define currents $j_{\pm }^{\star (k)}$ for $k=0,1,\dots
$ which are conserved in noncommutative space such that
\[
\partial _{-}j_{+}^{\star (k)}+\partial _{+}j_{-}^{\star (k)}=0,\qquad
\Leftrightarrow \qquad j_{\pm }^{\star (k)}=\pm \partial _{\pm }v^{(k)}.
\]
An infinite sequence of conserved non-local currents can be
obtained by iteration \footnote{ The non-local conserved currents
for the noncommutative models can also be constructed by using
bi-differential calculi and Hodge decomposition of the
differential forms for elements in the algebra of the
noncommutative torus \cite{dab1}-\cite{Muller3}. }
\[
j_{\pm }^{\star (k+1)}=D_{\pm }v^{(k)},\qquad \Rightarrow \qquad
\partial _{-}j_{+}^{\star (k+1)}+\partial _{+}j_{-}^{\star
(k+1)}=0.
\]
This establishes the equivalence of noncommutative Lax formalism
and noncommutative iterative construction. Here we have been able
to use nc-Lax formalism of nc-PCM to generate an infinite sequence
of non-local conserved quantities and have been able to relate
them with the nc-iterative procedure.

Let us make few comments about the algebra of these conserved
quantities. In the commutative case, the local conserved
quantities based on the invariant tensors, all Poisson commute
with each other and with the non-local conserved quantities. The
classical Poisson brackets of non-local conserved quantities
constitute classical Yangian symmetry $Y(u(N))$\cite{Hassan1}. The
Yangian is related to the Yang-Baxter equation \cite{new1} showing
the consistency with the factorization of multiparticle
$S$-matrix. The fundamental irreducible representations of the
Yangian correspond to particle multiplets and tensor product rules
of Yangian correspond to the Dorey's fusing rules \cite{new2}. In
the noncommutative case, we expect that noncommutative Yangian
appears in the model and its quantum version can shed some light
on the nonperturbative behavior of the model. One way of
investigating the algebra of non-local conserved quantities is to
develop a canonical formalism in noncommutative space and to
derive noncommutative Poisson bracket current algebra of the
model. In this work we have not attempted to answer these
questions and will return to these issues is some later work.

\section{Darboux Transformation of nc-PCM}\label{DTT}
The Lax pair of nc-PCM obtained in the previous section can be
further use to define Darboux transformation of generating
solutions of the linear system (\ref{l3}) of nc-PCM. We follow the
procedure of constructing Darboux transformation of PCM adopted in
\cite{DTPCM}. For convenience, we write the linear system
(\ref{l3}) can also be written as
\begin{eqnarray}
\partial _{+} v(x^{+},x^{-},\mu ) &=&\mu (2\mu -1)^{-1}j_{+}^{\star
}\star v (x^{+},x^{-},\mu ),  \label{Dlinear} \\
\partial _{-} v (x^{+},x^{-},\mu ) &=&\mu j_{-}^{\star }\star v
(x^{+},x^{-},\mu ),  \nonumber
\end{eqnarray}%
where $\mu=\frac{1-\gamma}{2}$ and $v(x^{+},x^{-},\mu )$ is a
non-degenerate $N\times N$ fundamental matrix solution of the
system (\ref{Dlinear}). The currents $j_{+}^{\star }$ and
$j_{-}^{\star }$ obey the following condition \footnote{The $U(N)$
group is composed of all $N\times N$ matrices. Then for all $g\in
U(N)$, $g^{\dagger}=g^{-1}$ where $g^{\dagger}$ is the hermitian
conjugate of $g$. A matrix $P\in u(N)$ Lie algebra of $U(N)$ if
and only if $P^{\dagger}+P=0$.}
\begin{eqnarray}
j_{+}^{\star }+j_{-}^{\star \dagger}=0.\label{DT0}
\end{eqnarray}
The equations (\ref{m1}) and (\ref{zero1}) can be written as
the compatibility condition of the linear system (\ref{Dlinear}) that is%
\[
\mu \left( \partial _{-}j_{+}^{\star }-\partial _{+}j_{-}^{\star
}+\left[ j_{+}^{\star },j_{-}^{\star }\right] _{\star }\right)
+(\mu -1)\left(
\partial _{-}j_{+}^{\star }+\partial _{+}j_{-}^{\star }\right) =0.
\]
In order to construct a noncommutative version of Darboux
transformation, we proceed as follows. $v[1]$ be another matrix
solution of the linear system (\ref{Dlinear}). The one-fold
Darboux transformation relates the solutions $v[1]$ and $v $ by
the following equation
\begin{eqnarray}
v [1]=D(\mu)\star v, \label{DT1}
\end{eqnarray}
where $D(\mu)$
\begin{eqnarray}
D(\mu)=I-\mu S,\label{DT2}
\end{eqnarray}
is the Darboux matrix and $S(x^{+},x^{-})$ is an $N\times N $ the
matrix function and $I$ is the identity matrix. The linear system
for the $v [1]$ is given by
\begin{eqnarray}
\partial _{+}v [1] &=&\mu (2\mu -1)^{-1}j_{+}^{\star
}[1]\star v [1],  \label{Dlinear1} \\
\partial _{-}v[1]&=&\mu j_{-}^{\star }[1]\star v
[1],  \nonumber
\end{eqnarray}
where $j_{+}^{\star }[1]$ and $j_{-}^{\star }[1]$ satisfy the
equations (\ref{m1}) and (\ref{zero1}). Applying $\partial_{\pm}$
on equation (\ref{DT1}) and equating the coefficients of different
powers of $\mu$, we get the following equations
\begin{eqnarray}
j_{+}^{\star }[1]=j_{+}^{\star }+\partial_{+}S, \qquad
j_{-}^{\star }[1]=j_{-}^{\star }-\partial_{-}S,\label{DT3}
\end{eqnarray}
and
\begin{eqnarray}
\partial_{+}S\star S-2\partial_{+}S &\equiv&S\star j_{+}^{\star
}-j_{+}^{\star
}\star S=[S,j_{+}^{\star }]_{\star}, \label{DT4}\\
\partial_{-}S\star S& \equiv&j_{-}^{\star }\star S-S\star
j_{-}^{\star }=-[S, j_{-}^{\star }]_{\star}.\nonumber
\end{eqnarray}
One can solve equation (\ref{DT4}) to get $S(x^{+},x^{-})$ so that
$j_{+}^{\star}[1]$, $j_{-}^{\star }[1]$ and $v[1]$ are obtained
from (\ref{DT3}), (\ref{DT1}) and (\ref{DT2}) respectively. An
explicit expression for the matrix $S(x^{+},x^{-})$ can be found
as follows.

Let us take $N$ complex numbers $\mu_{1},
\mu_{2},\dots,\mu_{N}(\neq 0,1/2)$ which are not all same. Also
take $N$ constant column vectors $w_{1}, w_{2}, \dots, w_{N}$ and
construct a non-degenerate $N\times N$ matrix
\begin{eqnarray}
M=\left(v(\mu_{1})w_{1},v(\mu_{2})w_{2},\dots,v(\mu_{N})w_{N}\right),\label{DT5}
\end{eqnarray}
with $\mbox{det}M\neq0$. Each column
$m_{\alpha}=v(\mu_{\alpha})w_{\alpha}$ in the matrix $M$ is a
solution of the linear system (\ref{Dlinear}) for
$\mu=\mu_{\alpha}$ i.e.
\begin{eqnarray}
\partial_{+}m_{\alpha}=\mu_{\alpha}(2\mu_{\alpha}-1)^{-1}j_{+}^{\star}\star
m_{\alpha},\qquad
\partial_{-}m_{\alpha}=\mu_{\alpha}j_{-}^{\star}\star
m_{\alpha},\label{DT9}
\end{eqnarray}
where $\alpha=1,2,\dots,N$. The matrix form of the equations
(\ref{DT9}) will be
\begin{eqnarray}
\partial_{+}M=j_{+}^{\star}\star
M \Lambda (2\Lambda-1)^{-1}, \qquad
\partial_{-}M=j_{-}^{\star}\star M \Lambda.\label{DT10}
\end{eqnarray}
 Let us take the matrix
\begin{eqnarray}
\Lambda=\mbox{diag}(\mu_{1}, \mu_{2},\dots,\mu_{N}).\label{DT6}
\end{eqnarray}
Such that the matrix
\begin{eqnarray}
S=M\star \Lambda^{-1} \star M^{-1},\label{DT7}
\end{eqnarray}
satisfies the equation (\ref{DT4}).

Our next step is to check that the equation (\ref{DT7}) is a
solution of the equation (\ref{DT4}). In order to show this, we
first apply $\partial_{+}$ on equation (\ref{DT7}) to get
\[
\partial_{+}S=(j_{+}^{\star}\star S-S\star j_{+}^{\star})\star M \star \Lambda (2\Lambda-1)^{-1}
\star M^{-1},
\]
or
\[
\partial_{+}S\star S-2\partial_{+}S\equiv S\star j_{+}^{\star}-j_{+}^{\star}\star S=[S,j_{+}^{\star}]_{\star}.
\]
Similarly, we apply $\partial_{-}$ on equation (\ref{DT7}) to get
\[
\partial_{-}S=j_{-}^{\star}-S\star j_{-}^{\star}\star S^{-1},
\]
or
\[
\partial_{-}S \star S\equiv j_{-}^{\star}\star S-S\star j_{-}^{\star}=-[S,j_{-}^{\star}]_{\star}.
\]
This shows that the equation (\ref{DT7}) is a solution of the
equation (\ref{DT4}). Equations (\ref{DT1}), (\ref{DT2}) and
(\ref{DT3}) define a Darboux transformation for the nc-PCM. In
order to have $j_{+}^{\star}[1], j_{-}^{\star}[1]\in u(N)$ we need
to show that
\begin{eqnarray}
\partial_{\pm}(S-S^{\dagger})=0,\label{DT13}
\end{eqnarray}
In other words we want to make specific $S$ to satisfies
(\ref{DT13}). This can be achieved if we choose
\begin{eqnarray}
\mu_{\alpha}=\left\{
\begin{array}{c}
\rho_{1} \,\,\,(\alpha=1,2,\dots,k)\\(0<k<N)\\
\rho_{2}\,\,\,\,(\alpha=k+1,k+2,\dots,N)\end{array}%
\right.\label{DT11}
\end{eqnarray}
Now take $\rho_{1}$ to be an imaginary number and define
\begin{eqnarray}
\rho_{2}=\frac{\bar{\rho}_{1}}{2\bar{\rho}_{1}-1},\label{p}\end{eqnarray}
with $|2\rho_{1}-1|\neq1$ so that $\rho_{1}\neq\rho_{2}$. This has
been defined for the later convenience. Now we define column
solutions for eigenvalues $\rho_{1}$ and $\rho_{2}$. Let
$m_{1},m_{2},\dots,m_{k}$ and $m_{k+1},m_{k+2},\dots,m_{N}$ be
column solution of the linear system (\ref{Dlinear}) for
$\mu=\rho_{1}$ and $\mu=\rho_{2}$ respectively i.e.
\begin{eqnarray}
m_{p}=v(\rho_{1})w_{p},\qquad m_{q}=v(\rho_{2})w_{q}.\label{DT12}\\
(p=1,2,\dots,k,\qquad q=k+1,k+2,\dots,N)\nonumber
\end{eqnarray}
We have to choose $w_{\alpha}$ so that
\begin{eqnarray}
m^{\dagger}_{q}\star m_{p}=0,\qquad (p=1,2,\dots,k,\qquad
q=k+1,k+2,\dots,N).\label{DT14}
\end{eqnarray}
at one point (say $(0,0)$) and $m_{\alpha}$ are linearly
independent. We shall show that the matrix $S$ constructed from
these values of $\mu_{\alpha}$ and $w_{\alpha}$ satisfies
(\ref{DT13}).

First of all, we prove that (\ref{DT14}) holds everywhere if it
holds at one point. The proof of the above identity (\ref{DT14})
is as follows. Let us first calculate
\begin{eqnarray}
\partial_{-}(m^{\dagger}_{q}\star
m_{p})&=&\partial_{-}m^{\dagger}_{q}\star
m_{p}+m^{\dagger}_{q}\star\partial_{-}
m_{p}\nonumber\\&=&\left(\partial_{+}m_{q}\right)^{\dagger}\star
m_{p}+m^{\dagger}_{q}\star\partial_{-}
m_{p},\nonumber\\&=&\left(\frac{\rho_{2}}{2\rho_{2}-1}j_{+}^{\star}\star
m_{q} \right)^{\dagger}\star
m_{p}+m^{\dagger}_{q}\star\partial_{-}
m_{p},\nonumber\\&=&-\left(\frac{\bar{\rho}_{2}}{2\bar{\rho}_{2}-1}\right)m_{q}
^{\dagger}\star j_{-}^{\star}\star
m_{p}+m^{\dagger}_{q}\star\partial_{-}
m_{p},\nonumber\\&=&-\rho_{1}m_{q} ^{\dagger}\star
j_{-}^{\star}\star m_{p}+\rho_{1}m_{q} ^{\dagger}\star
j_{-}^{\star}\star m_{p},\nonumber\\&=&0.\nonumber
\end{eqnarray}
Similarly we can check
\begin{eqnarray}
\partial_{+}(m^{\dagger}_{q}\star m_{p})=0.\nonumber
\end{eqnarray}
This means that (\ref{DT14}) holds everywhere if it holds at one
point i.e.
\begin{eqnarray}
\partial_{\pm}(m^{\dagger}_{q}\star m_{p})=0.\label{DT16}
\end{eqnarray}
$m_{\alpha}'s$ are linearly independent everywhere if they are
linearly independent at one point as (\ref{DT9}) is linear. We
have to choose $m_{\alpha}$ so that they are linearly independent
and (\ref{DT16}) holds everywhere.

From equation (\ref{DT7}) we have
\begin{eqnarray}
S\star m_{p}=\frac{1}{\rho_{1}}m_{p},\qquad S\star
m_{q}=\frac{1}{\rho_{2}}m_{q}.\label{DT18}
\end{eqnarray}
The hermitian conjugate of equation (\ref{DT18}) is given by
\begin{eqnarray}
m^{\dagger}_{p}\star
S^{\dagger}=\frac{1}{\bar{\rho}_{1}}m^{\dagger}_{p},\qquad
m^{\dagger}_{q}\star
S^{\dagger}=\frac{1}{\bar{\rho}_{2}}m^{\dagger}_{q}.\label{DT19}
\end{eqnarray}
Therefore
\begin{eqnarray}
m^{\dagger}_{p}\star(S^{\dagger}-S)\star m_{r}
=(\frac{1}{\bar{\rho}_{1}}-\frac{1}{\rho_{1}})m^{\dagger}_{p}
\star m_{r},\label{DT20}\\
m^{\dagger}_{q}\star(S^{\dagger}-S)\star
m_{s}=(\frac{1}{\bar{\rho}_{2}}-\frac{1}{\rho_{2}})m^{\dagger}_{q}
\star m_{s}.\nonumber
\end{eqnarray}
or
\begin{eqnarray}
&&\left.m^{\dagger}_{q}\star(S^{\dagger}-S)\star m_{p}
=0,\right.\label{DT21}\\
&&\left.m^{\dagger}_{p}\star(S^{\dagger}-S)\star m_{q}=0.\right.\nonumber \\
&&\left.\qquad(p,r=1,2,\dots,k, \qquad
q,s=k+1,k+2,\dots,N)\right.\nonumber
\end{eqnarray}
From equation (\ref{p}) we have
\begin{eqnarray}
\frac{1}{\rho_{1}}-\frac{1}{\bar{\rho}_{1}}=\frac{1}{\rho_{2}}-\frac{1}{\bar{\rho}_{2}}.\label{DT22}
\end{eqnarray}
Take
\begin{eqnarray}
m^{\dagger}_{\beta}\star(S^{\dagger}-S)\star m_{\alpha}
=m^{\dagger}_{\beta}\star (\frac{1}{\bar{\rho}_{1}}-\frac{1}{\rho_{1}})\star I\star m_{\alpha},\label{DT23}\\
(\alpha, \beta=1,2,\dots,N)\nonumber
\end{eqnarray}
and $\frac{1}{\rho_{1}}-\frac{1}{\rho_{2}}$ is real. Since the set
$\{m_{\alpha}\}$ consists of $N$ linearly independent vectors
\begin{eqnarray}
S^{\dagger}-S
=(\frac{1}{\bar{\rho}_{1}}-\frac{1}{\rho_{1}})I,\label{DT24}
\end{eqnarray}
therefore the equation (\ref{DT24}) implies that
\begin{eqnarray}
j^{\star}_{+}[1]+j^{\star \dagger }_{-}[1]=j^{\star}_{+}+j^{\star
\dagger }_{-}-\partial_{+}(S^{\dagger}-S)=0.\label{DT25}
\end{eqnarray}
This proves that $j^{\star}_{+}[1]$ and $j^{\star \dagger}_{-}[1]$
satisfy the equation (\ref{DT0}) for $U(N)$. To summarize our
results, we write the one-fold Darboux transformation as
\begin{eqnarray*}
(j^{\star}_{+}, j^{\star}_{-}, v)\longrightarrow
(j^{\star}_{+}[1], j^{\star}_{-}[1], v[1]), \end{eqnarray*} where
\begin{eqnarray*}
&&\left.v[1]=(I-\mu S)v,\right.\\
&&\left.j^{\star}_{+}[1]=j^{\star}_{+}+\partial_{+}S, \qquad
j^{\star}_{-}[1]=j^{\star}_{-}-\partial_{-}S,\right.
\end{eqnarray*}
and $v[1]$ is the solution of the following linear system
\begin{eqnarray*}
\partial _{+}v [1] =\mu (2\mu -1)^{-1}j_{+}^{\star
}[1]\star v [1],  \qquad \partial _{-}v[1]=\mu j_{-}^{\star
}[1]\star v [1],
\end{eqnarray*}
such that the matrix $S$ satisfies the following equations
\begin{eqnarray*}
\partial_{+}S\star S-2\partial_{+}S=[S,j_{+}^{\star
}]_{\star},\qquad \partial_{-}S\star S=-[S, j_{-}^{\star
}]_{\star},\nonumber
\end{eqnarray*}
and the currents $j_{+}^{\star}[1], j_{-}^{\star}[1]$ satisfy the
following condition
\begin{eqnarray*}
j^{\star}_{+}[1]+j^{\star \dagger}_{-}[1]=0.
\end{eqnarray*}
The two-fold Darboux transformation is
\begin{eqnarray*}
(j^{\star}_{+}[1], j^{\star}_{-}[1], v[1])\longrightarrow
(j^{\star}_{+}[2], j^{\star}_{-}[2], v[2]), \end{eqnarray*}
 where
\begin{eqnarray*}
&&\left.v[2]=(I-\mu S[1])v[1],\right.\\
&&\left.j^{\star}_{+}[2]=j^{\star}_{+}[1]+\partial_{+}S[1], \qquad
j^{\star}_{-}[2]=j^{\star}_{-}[1]-\partial_{-}S[1].\right.
\end{eqnarray*}
and $v[2]$ is the solution of the following linear system
\begin{eqnarray*}
\partial _{+}v [2] =\mu (2\mu -1)^{-1}j_{+}^{\star
}[2]\star v [2],  \qquad \partial _{-}v[2]=\mu j_{-}^{\star
}[2]\star v [2],
\end{eqnarray*}
such that the matrix $S[1]$ satisfies the following equations
\begin{eqnarray*}
\partial_{+}S[1]\star S[1]-2\partial_{+}S[1]=[S[1],j_{+}^{\star
}[1]]_{\star},\qquad \partial_{-}S[1]\star S[1]=-[S[1],
j_{-}^{\star }[1]]_{\star},\nonumber
\end{eqnarray*}
and the matrix $S[1]$ is given by
\begin{eqnarray*}
S[1]=M[1]\star \Lambda^{-1} \star M[1]^{-1},\end{eqnarray*}with
$M[1]$ obeying
\begin{eqnarray*}
\partial_{+}M[1]=j_{+}^{\star}[1]\star M[1]
\Lambda(2\Lambda-1)^{-1},\qquad
\partial_{-}M[1]=j_{-}^{\star}[1]\star M[1] \Lambda,
\end{eqnarray*}
and the currents $j_{+}^{\star}[2], j_{-}^{\star}[2]$ satisfy the
following condition
\begin{eqnarray*}
j^{\star}_{+}[2]+j^{\star \dagger}_{-}[2]=0.
\end{eqnarray*}

The result can be generalized to obtain $K$-fold Darboux
transformation
\begin{eqnarray*}
(j^{\star}_{+}[K-1], j^{\star}_{-}[K-1], v[K-1])\longrightarrow
(j^{\star}_{+}[K], j^{\star}_{-}[K], v[K]), \end{eqnarray*}
 where
\begin{eqnarray*}
&&\left.v[K]=(I-\mu S[K-1])v[K-1],\right.\\
&&\left.j^{\star}_{+}[K]=j^{\star}_{+}[K-1]+\partial_{+}S[K-1],
\qquad
j^{\star}_{-}[K]=j^{\star}_{-}[K-1]-\partial_{-}S[K-1],\right.
\end{eqnarray*}
and $v[K]$ is the solution of
\begin{eqnarray*}
\partial _{+}v [K] =\mu (2\mu -1)^{-1}j_{+}^{\star
}[K]\star v [K],  \qquad \partial _{-}v[K]=\mu j_{-}^{\star
}[K]\star v [K],
\end{eqnarray*}
such that the matrix $S[K-1]$ satisfies the following equations
\begin{eqnarray*}
\partial_{+}S[K-1]\star S[K-1]-2\partial_{+}S[K-1]=[S[K-1],j_{+}^{\star
}[K-1]]_{\star},\\ \partial_{-}S[K-1]\star S[K-1]=-[S[K-1],
j_{-}^{\star }[K-1]]_{\star}.\nonumber
\end{eqnarray*}
The matrix $S[K-1]$ is given by
\begin{eqnarray*}
S[K-1]=M[K-1]\star \Lambda^{-1} \star M[K-1]^{-1},\end{eqnarray*}
such that $M[K-1]$ obeys
\begin{eqnarray*}
\partial_{+}M[K-1]=j_{+}^{\star}[K-1]\star M[K-1]
\Lambda(2\Lambda-1)^{-1},\qquad
\partial_{-}M[K-1]=j_{-}^{\star}[K-1]\star M[K-1] \Lambda.
\end{eqnarray*}
where the currents $j_{+}^{\star}[K], j_{-}^{\star}[K]$ satisfy
the following condition
\begin{eqnarray*}
j^{\star}_{+}[K]+j^{\star \dagger}_{-}[K]=0.
\end{eqnarray*}
This completes the iteration of Darboux transformation. From a
given seed solution one can generate the noncommutative
multi-soliton solutions of the system. Such solutions have been
constructed for the noncommutative integrable $U(N)$ sigma model
in 2+1 dimensions by employing the dressing method and their
scattering properties have been investigated
\cite{new11}-\cite{new22}. In addition, these multi-soliton
solutions correspond to D0-branes moving inside the D2-branes in
open $N=2$ fermionic string theory \cite{new33}-\cite{new44}. For
two dimensional Euclidean sigma models, the noncommutative
multi-solitons and their moduli-space have been constructed that
unifies different descriptions of abelian and non-abelian
multi-solitons \cite{new55}. For nc-PCM such solutions can be
explicitly constructed either by Darboux transformation or by
dressing method. One can also generalize the uniton method
\cite{beck}-\cite{wood} of constructing solutions of nc-PCM. These
non-trivial solutions of nc-PCM are presented for a given
projection operator in \cite{Lee} where the construction of the
unitons for nc-PCM is based on the noncommutative generalization
of the theorem due to K. Uhlenbeck \cite{beck}-\cite{wood}. We
shall return to the complete description of construction of uniton
solutions of nc-PCM and proof of noncommutative version of the
theorem of K. Uhlenbeck in some later work.

\section{Perturbative Expansion}\label{perexp}
In this section we study the perturbative expansion of the
noncommutative fields of nc-PCM and compute the equation of motion
and the conserved quantities upto first order in perturbative
expansion in noncommutativity parameter $\theta$. We can expand
the currents $j_{\pm }$ as power series in the $\theta$. We expand
the currents $j_{\pm }$ upto
first order in $\theta $%
\begin{equation}
j_{\pm }^{\star }=j_{\pm }^{[0]}+\theta \,\,j_{\pm }^{[1]}.
\label{current1}
\end{equation}
By substituting the value of $j_{\pm }^{\star }$ from equation
(\ref{current1}) in equation of motion (\ref{m1}) and
(\ref{zero1}), we get
\begin{eqnarray}
\partial _{-}j_{+}^{[0]}+\partial _{+}j_{-}^{[0]} &=&0,\nonumber \\
\partial _{-}j_{+}^{[1]}+\partial _{+}j_{-}^{[1]} &=&0,\nonumber\\
\partial _{-}j_{+}^{[0]}-\partial _{+}j_{-}^{[0]}+\left[
j_{+}^{[0]},j_{-}^{[0]}\right]  &=&0,\nonumber \\
\partial _{-}j_{+}^{[1]}-\partial _{+}j_{-}^{[1]}+\left[
j_{+}^{[1]},{j}_{-}^{[0]}\right] +\left[ {j}_{+}^{[0]},j%
_{-}^{[1]}\right]  &=&-\frac{\mbox{i}}{2}\left(
j_{++}^{[0]}j_{--}^{[0]}+j_{--}^{[0]}j_{++}^{[0]}\right)
-\frac{\mbox{i}}{8}\left[
j_{+}^{[0]},j_{-}^{[0]}\right]^2.\label{per11}
\end{eqnarray}
where $j_{\pm\pm}^{[0]}=\partial_{\pm}j_{\pm}^{[0]}$. It is clear
from the above equations that the currents $j_{\pm}^{[0]}$ and
$j_{\pm}^{[1]}$ are conserved, $j_{\pm}^{[0]}$ is curvature free
but $j_{\pm}^{[1]}$ is not curvature free.

The perturbative expansion of iterative construction gives the
following results
\begin{eqnarray*}
j_{\pm }^{[0](k+1)}&=&D_{\pm
}^{[0]}v^{[0](k)},\,\,\,\,\,\,\,\qquad\qquad\qquad \Rightarrow
\qquad\partial _{-}j_{+}^{[0](k+1)}+\partial
_{+}j_{-}^{[0](k+1)}=0,
\\j_{\pm }^{[1](k+1)}&=&D_{\pm
}^{[0]}v^{[1](k)}-D_{\pm }^{[1]}v^{[0](k)},\qquad \Rightarrow
\qquad
\partial _{-}j_{+}^{[1](k+1)}+\partial
_{+}j_{-}^{[1](k+1)}=0,
\end{eqnarray*}
where
\begin{eqnarray*}
D_{\pm}^{[0]}v^{[0](k)}&=&\partial_{\pm}v^{[0](k)}-j_{\pm}^{[0]}v^{[0](k)},\\
D_{\pm}^{[0]}v^{[1](k)}&=&\partial_{\pm}v^{[1](k)}-j_{\pm}^{[0]}v^{[1](k)},\\
D_{\pm}^{[1]}v^{[0](k)}&=&j_{\pm}^{[1]}v^{[0](k)}+\frac{\mbox{i}}{2}%
(\partial_{\pm}j_{\pm}^{[0]}\partial_{\mp}-\partial_{\mp}j_{\pm}^{[0]}\partial_{\pm})v^{[0](k)}.
\end{eqnarray*}
From this analysis, it is obvious that the conservation of $k$th
current implies the conservation of $(k+1)$th current at zeroth as
well as first order of perturbation expansion in the parameter of
noncommutativity.

By substituting the value of $j_{\pm }^{\star }$ from equation
(\ref{current1}) in equation (\ref{localseries}), we obtain first
four zeroth and first order local conserved quantities
\begin{eqnarray*}
&&\left. \partial _{\mp }{\rm Tr}\left( j_{\pm }^{[0]2}\right)
=0,\right.  \\
&&\left. \partial _{\mp }{\rm Tr}\left( j_{\pm
}^{[0]}j_{\pm}^{[1]}\right)
=0,\right.  \\
&&\left. \partial _{\mp }{\rm Tr}\left( j_{\pm }^{[0]3}\right)
=0,\right.  \\
&&\left. \partial _{\mp }{\rm Tr}\left( j_{\pm
}^{[0]2}j_{\pm}^{[1]}-\frac{1}{4}j_{\pm }^{[0]}j_{\pm \pm }^{[0]}\left[ j%
_{\pm }^{[0]},j_{\mp }^{[0]}\right]+\frac{1}{4}j_{\pm\pm
}^{[0]}j_{\pm}^{[0]}[%
j_{\pm }^{[0]},j_{\mp}^{[0]}]\right) =0,\right.  \\
&&\left. \partial _{\mp }{\rm Tr}\left( j_{\pm }^{[0]4}\right)
=0,\right.  \\
&&\left. \partial _{\mp }{\rm Tr}\left( j_{\pm
}^{[0]3}j_{\pm}^{[1]}-\frac{1}{8}j_{\pm }^{[0]2}j_{\pm \pm
}^{[0]}\left[ j_{\pm }^{[0]},j_{\mp
}^{[0]}\right]+\frac{1}{8}j_{\pm\pm }^{[0]}j_{\pm }^{[0]2}\left[
j_{\pm }^{[0]},j_{\mp }^{[0]}\right] \right) =0.\right.
\end{eqnarray*}
The conservation laws hold because of the equation (\ref{per11}).
The conserved holomorphic currents $\partial _{\mp }{\rm Tr}\left(
j_{\pm }^{[0]}\right)^2, \partial _{\mp }{\rm Tr}\left( j_{\pm
}^{[0]}\right)^3, \partial _{\mp }{\rm Tr}\left( j_{\pm
}^{[0]}\right)^4,\dots$ are the usual local currents and the
corresponding local conserved quantities are
\[
Q^{[0]}_{\pm s}=\int_{-\infty }^{\infty }dx{\rm Tr}\left( j_{\pm
}^{[0]}\right)^n,
\]
where $s=n-1$ represents the spin of the conserved quantity. The
higher spin conserved quantities are in involution with each other
i.e. \begin{eqnarray*}
\{Q^{[0]}_{+ s}, Q^{[0]}_{- r} \}&=&0, \qquad r,s>0\\
\{Q^{[0]}_{\pm s},Q^{[0]}_{\pm r}\}&=&0.
\end{eqnarray*}
The values of $s$ are precisely the exponents modulo the Coxeter
number of Lie algebra $u(N)$. These conservation laws are also
related to the symmetric invariant tensors of the $u(N)$ and the
zeroth order contributions give the commuting conserved quantities
with spins equal to the exponents of the underlying algebra. We
also expect that the first order conserved quantities are also in
involution and the calculations involve the Poisson bracket
current algebra of the model which we have not been able to find
at this stage.

Similarly we can expand conserved quantities
\begin{eqnarray*}
Q^{(1)[0]} &=&-\int_{-\infty }^{\infty }j_{0}^{[0]}(x^{0},y)dy, \\
Q^{(1)[1]} &=&-\int_{-\infty }^{\infty }j_{0}^{[1]}(x^{0},y)dy, \\
Q^{(2)[0]} &=&\int_{-\infty }^{\infty }\left(
-j_{1}^{[0]}(x^{0},y)+j_{0}^{[0]}(x^{0},y)\int_{-\infty
}^{y}j_{0}^{[0]}(x^{0},z)dz\right) dy, \\
Q^{(2)[1]} &=&\int_{-\infty }^{\infty }\left( -j%
_{1}^{[1]}(x^{0},y)+j_{0}^{[1]}(x^{0},y)\int_{-\infty
}^{y}j_{0}^{[0]}(x^{0},z)dz+j_{0}^{[0]}(x^{0},y)\int_{-\infty }^{y}j%
_{0}^{[1]}(x^{0},z)dz\right) dy.
\end{eqnarray*}
The conservation of these quantities can be proved by using
equation (\ref{per11}). The non-local conserved quantities
$Q^{(1)[0]}$ and $Q^{(2)[0]}$ form the usual Yangian $Y(u(N))$.
There are two copies of this structure corresponding to right and
left currents and therefore the algebra is $Y_{L}(u(N))\times
Y_{R}(u(N))$. These zeroth order Yangian conserved quantities also
Poisson commutate with the zeroth order local conserved quantities
$Q^{[0]}_{\pm s}$ i.e.
 \begin{eqnarray*}
\{Q^{[0]}_{\pm s}, Q^{(1)[0]} \}&=&0,\\
\{Q^{[0]}_{\pm s}, Q^{(2)[0]} \}&=&0.
\end{eqnarray*}
The first order contribution in the algebra of both local and
non-local conserved quantities can be investigated if the deformed
canonical Poisson bracket algebra of deformed currents is known.
Note that the first order correction to the first non-local
conserved quantity is an integral of non-local function of the
fields. These corrections shall naturally modify the Yangian
structure of the non-local conserved quantities and as a result a
Moyal deformed Yangian might appear, whose zeroth order element
must be the usual Yangian of the given Lie algebra.
\section{Conclusions}\label{remarks}
In this paper, we have analyzed the Lax formalism of nc-PCM. In
this generalization, we have observed that noncommutative
extension works straightforwardly resulting in a noncommutative
equation of PCM without any constraint appearing due to
noncommutativity. The Lax formalism of nc-PCM has been used to
generate local as well as non-local conserved quantities of the
model and it has been shown that the Lax formalism of nc-PCM is
equivalent to the iterative procedure already used in
\cite{Marco1}. Furthermore, the Lax formulism has been used to
derive $K$-fold Darboux transformation of the nc-PCM. The
noncommutative Darboux transformation can be used to construct
non-trivial solutions of the nc-PCM and to study their
moduli-space dynamics. The present work can be extended to
construct the uniton solutions of nc-PCM and to investigate the
algebra of local and non-local conserved quantities. Another
interesting direction to pursue is to look at the quantum
conservation of the local and non-local conserved quantities of
the nc-PCM. The method of anomaly counting for the quantum
mechanical survival of the local conservation laws can also be
applied to nc-PCM \cite{Hassan1}-\cite{new2}. For the non-local
conserved quantities the quantum Yangians can also be investigated
for the nc-PCM \cite{Luc}-\cite{Bernard}. It is also interesting
to seek noncommutative extension of the Lax formalism, local and
non-local conserved quantities of supersymmetric PCM in the
direction adopted in \cite{Hassan2}-\cite{saleem} for the
commutative supersymmetric PCM.
\bigskip

{\large {\bf {Acknowledgements}}}

We acknowledge the enabling role of the Higher Education
Commission, Pakistan and appreciate its financial support through
``Merit Scholarship Scheme for PhD studies in Science \&
Technology (200 Scholarships)''.


\begin{thebibliography}{99}

\bibitem{N1}  S. Minwalla, M. V. Raamsdonk and N. Seiberg, J. High Energy Phys. {\bf 02} (2000) 020.

\bibitem{N2}  N. Seiberg and E. Witten, J. High Energy Phys. {\bf 09} (1999) 032.


\bibitem{Fur}  K. Furuta and T. Inami, Mod. Phys. Lett. \textbf{A15} (2000)
997.

\bibitem{Marco1}  M. Moriconi and I.C. Carnero, Nucl. Phys. {\bf B673} (2003) 437.


\bibitem{Penati1}  M.T. Grisaru and S. Penati, Nucl.Phys. {\bf B655} (2003) 250.


\bibitem{Hamanaka1}  M. Hamanaka,  J. Math. Phys. \textbf{46}(2005) 052701.

\bibitem{Hamanaka3}  M. Hamanaka and K. Toda, Phys. Lett. \textbf{A316} (2003) 77.

\bibitem{dab1} L. Dabrowski, T. Krajewski and G. Landi, Mod. Phys. Lett. \textbf{A18} (2003) 2371.


\bibitem{Muller1}  A. Dimakis and F. M. Hoissen, Int. J. Mod. Phys. \textbf{B14} (2000) 2455.


\bibitem{Muller3}  A. Dimakis and F. M. Hoissen,  J. Phys. \textbf{A37} (2004) 4069.



\bibitem{usman1}  U. Saleem, M. Hassan and M. Siddiq, J. Phys. \textbf{A38} (2005) 9241.

\bibitem{usman}  U. Saleem, M. Siddiq and M. Hassan, Chin. Phys. Lett. \textbf{22} (2005) 1076.


\bibitem{Takasaki}K. Takasaki, J. Geom. Phys. \textbf{37} (2001) 291.

\bibitem{se}  N. Seiberg, L. Susskind and N. Toumbas, J. High Energy Phys. \textbf{06} (2000) 044.

\bibitem{go}  J. Gomis and T. Mehen, Nucl. Phys. {\bf B591} (2000) 265.

\bibitem{moh}  I. Jack, D. R. T. Jones and N. Mohammedi, Phy. Lett. {\bf B520} (2001) 405.

\bibitem{moyal}  J. E. Moyal, Proc. Cambridge. Phil. Soc. {\bf 45} (1949) 99.

\bibitem{BIZZ}  E. Br\'{e}zin, C. Itzykson, J. Zinn-Justin and J.-B. Zuber, Phys. Lett. {\bf B82 }(1979) 442.


\bibitem{R1} K. Matsubara, Phys. Lett. \textbf{B482} (2000) 417.

\bibitem{R2} L. Bonora, M. Schnable, M. M. Sheikh-Jabbari and
A. Tomasiell, Nucl. Phys. \textbf{B589} (2000) 461.

\bibitem{R3} I. Bars, M. M. Sheikh-Jabbari and
M. Vasiliev, Phys. Rev. \textbf{D64} (2000) 086004.

\bibitem{Gold} Y. Y. Goldschmidt and E. Witten, Phys. Lett. \textbf{B91} (1980) 392.

\bibitem{y1} A. B. Zamolodchikov and Al. B. Zamolodchikov, Ann. Phys. \textbf{120} (1979) 253.

\bibitem{y2} S. Parke, Nucl. Phys. {\bf B174 }(1980) 166.

\bibitem{y3} E. Ogievetsky, P. Wiegmann and N. Reshetikhin, Nucl. Phys. {\bf B280 }(1987) 45.

\bibitem{y4} E. Frenkel and N. Reshetikhin, Commun. Math. Phys. \textbf{197} (1998) 1.

\bibitem{Hassan1}  J. M. Evans, M. Hassan, N. J. MacKay and A. J. Mountain,
Nucl. Phys. {\bf B561} (1999) 385.\\J. M. Evans, D. Kagan, N. J.
MacKay, C. A. S. Young, J. High Energy Phys. \textbf{01} (2005)
020.

\bibitem{new1}M. Jimbo, Int. J. Mod. Phys. \textbf{A4} (1989) 3759.

\bibitem{new2}P. Dorey, Nucl. Phys. \textbf{B358} (1991) 654;\\
H. Braden, J. Phys. \textbf{A25} (1992) L15;\\
V. Chari and A. Pressley, Commun. Math. Phys. \textbf{181} (1996)
265.


\bibitem{DTPCM} C. Gu, H. Hu and Z. Zhou, {\it Darboux
transformations in integrable systems: theory and their
applications to geometry}, Springer-Verlag (2005).


\bibitem{new11} O. Lechtenfeld and A. D. Popov, J. High Energy Phys. \textbf{11} (2001) 040 .


\bibitem{new22}O. Lechtenfeld and A. D. Popov, Phys. Lett. {\bf B523} (2001) 178.

\bibitem{new33} O. Lechtenfeld, A. D. Popov and B. Spending, J. High Energy Phys. \textbf{06} (2001) 011.

\bibitem{mann} M. Ihl and S. Uhlmann, Int. J. Mod. Phys. {\bf A18}
(2003) 4889.
\bibitem{new44} M. Wolf, J. High Energy Phys. \textbf{06} (2002) 055.

\bibitem{new55} A. V. Domrin, O. Lechtenfeld and S. Petersen, J. High Energy Phys. \textbf{03} (2005) 045.


\bibitem{beck} K. Uhlenbeck, J. Diff. Geom. {\bf 30} (1989) 1.

\bibitem{wood} J. C. Wood, Proc. London. Math. Soc. {\bf 58}
(1989) 608.

\bibitem{Lee} K. M. Lee, J. High Energy Phys. \textbf{08} (2004) 054.

\bibitem{Luc} M. Luscher, Nucl. Phys. {\bf B135 }(1978) 1.

\bibitem{Bernard} D. Bernard and A. LeClair, Phys. Lett.
\textbf{B247} (1990) 309; Commun. Math. Phys. \textbf{142} (1991)
99; Nucl. Phys. \textbf{B399} (1993) 709; \\D. Bernard, Commun.
Math. Phys. \textbf{137} (1991) 191.

\bibitem{Hassan2}  J. M. Evans, M. Hassan, N. J. MacKay and A. J. Mountain,
Nucl. Phys. {\bf B580 }(2000) 605.

\bibitem{saleem}  U. Saleem and M. Hassan, Eur. Phys. J. {\bf C38} (2005) 521.
\end{thebibliography}
\end{document}